\documentclass[reprint,aps,prb,twocolumn,superscriptaddress,showpacs]{revtex4-1}
\usepackage{graphicx}
\usepackage{mathrsfs}
\usepackage{bm}
\usepackage{hyperref}
\usepackage{dcolumn}
\usepackage{amsmath}
\usepackage{amssymb}
\usepackage{color}
\usepackage{subfig}
\usepackage{caption}
\hypersetup{  colorlinks=true, linkcolor=blue, citecolor=red, urlcolor=blue  }

\newcommand{\bk}{{\mathbf k}}
\newcommand{\bp}{{\mathbf p}}

\begin{document}

\title{Electronic properties of thin films of tensile strained HgTe}
\author{Jihang Zhu}
\affiliation{Department of Physics, The University of Texas at Austin, Austin, Texas 78712,USA}
\author{Chao Lei}
\affiliation{Department of Physics, The University of Texas at Austin, Austin, Texas 78712,USA}
\author{Allan H. MacDonald}
\affiliation{Department of Physics, The University of Texas at Austin, Austin, Texas 78712,USA}
\email{macd@physics.utexas.edu}

\begin{abstract}
Tensile strained bulk HgTe is a three-dimensional topological insulator.  
Because of the energetic position of 
its surface state Dirac points relative to its small bulk gap, 
the electronic properties of the relatively thin MBE-grown films 
used to study this material experimentally are quite sensitive to details of its electrostatic band-bending physics.
We have used an 8-band $\mathbf{k} \cdot \mathbf{p}$ model to 
evaluate the gate voltage dependence of its thin-film two-dimensional 
subbands and related thermodynamic and transport properties
in films with thicknesses between 30 and 70nm, accounting self-consistently 
for gate field screening by
the topologically protected surface states and bulk state response.
We comment on the effective dielectric constant $\epsilon$ 
that is appropriate in calculations of this type, arguing for a smaller value 
$\epsilon_r \approx 6.5$ than is commonly used.
Comparing with recent experiments, we find that our fully microscopic model of gate field screening
alters the interpretation of some observations that have been used to 
characterize strained HgTe thin films.  
\end{abstract}

\maketitle

\section{Introduction}
Three-dimensional(3D) topological insulators(TIs) possess 2D Dirac-like surface states that are protected by 
time-reversal symmetry\cite{RevTI2010,RevTI2011} and have interesting transport properties related
to strong spin-momentum locking.  Unintended and uncontrolled bulk doping often obscures 
the meaning of experiments that are intended to probe the surface state properties of the heavily studied 
pnictogen chalcogenide bulk TIs like $\text{Bi}_2\text{Se}_3, \text{Bi}_2\text{Te}_3$ and $\text{Sb}_2\text{Te}_3$, 
making it hard to distinguish surface states contributions to observables from bulk 
contributions.  Tensile strained HgTe is an alternative 3D topological insulator\cite{Fu2007} 
that has the advantage of extremely low background doping, which should make it more 
possible to separate surface and bulk effects experimentally. 
Considerable progress has already been made studying  
strained HgTe films experimentally, including realizations of proximitied superconductivity\cite{Wiedenmann2016,Hart2016,Bocquillon2016,Hart2014,Sochnikov2015}, 
the quantum Hall effect\cite{Strain2011} and the quantum anomalous Hall effect\cite{Budewitz2017}.

However, the strain is practically induced by lattice mismatch in quantum wells. That is, only thin films of strained HgTe can be made experimently. And the bulk gap of the strained HgTe thin films is proportional to the strength of tensile strain\cite{Strain2014}. Because of the small strain, the bulk gap is small as well. The bulk gap of the strained HgTe/CdTe quantum well is about $20$meV, while the gaps of pnictogen chalcogenides are about $0.3$eV. Because of the small gap, the surface state localization length is fairly large and the separation between bulk and surface effects in thin film HgTe samples that are available experimentally is not always clear. Efforts to access and verify the surface states of strained HgTe are still making\cite{Diracscreening,Qcapacitance2016}, mainly on thin films up to thickness of 80nm.

With these motivations, we have examined the multi-band envelope description of strained HgTe thin films with thicknesses between 30 and 70nm, accounting carefully for electron-electron interactions. Our model provides a basis to interpret electronic properties of surface and bulk states, tuned by varying gate voltage in experiments\cite{Diracscreening,Qcapacitance2016,Thermalpower2017}. We find that the films respond like semiconductors to gates that induce electron gases, but like metals to gates that induce hole gases.  We are able to provide simple explanations for recent experiments that characterize HgTe thin films using capacitance\cite{Qcapacitance2016} and thermopower measurements\cite{Thermalpower2017}.
In particular, we find that it is not necessary to invoke phonon drag to explain the large difference in thermopower between electron and hole cases.

The paper is organized as follows.
In section \ref{TheoreticalModel} we introduce the theoretical model that are used in our paper. 8-band $\bk \cdot \bp$ theory along with electron-electron interactions are used. The strain effect is considered by Bir-Pikus model.
In section \ref{DielectricConstant}, we discuss the dielectric constant of strained HgTe thin films and we find that it should be much smaller than the value, 21, which is widely used.
In section \ref{QuantumCapacitance}, we capture the capacitance of strained HgTe quantum wells based on our established model and directly compare results with recent capacitance experiments.
In section \ref{Thermoelectrics}, transport coefficient, thermopower and the Nernst coefficient

\section{$\bk \cdot \bp$ theory for strained $\text{HgTe}$}\label{TheoreticalModel}
HgTe has a zinc-blende crystal structure which lacks inversion symmetry. The crystalline HgTe is a semimetal with conduction band ($\Gamma_8$ light-hole) and valence band ($\Gamma_8$ heavy-hole) degenerate at the $\Gamma$ point.
The band structure of HgTe/CdTe quantum well in [001] direction is calculated by self-consistently solving the Poisson-Schr\"{o}dinger equation using the 8-band $\bm{k}\cdot \bm{p}$ approach\cite{Hamiltonian2005}:
\begin{equation}\label{Hamiltonian}
\begin{aligned}
\begin{split}
 &\sum\limits_{n^{\prime}} (H_{nn^{\prime}} +\phi_{\text{H}}\delta_{nn^{\prime}}) \psi_{n^{\prime}} = E \psi_{n}, \\
 &\nabla^2_z V_{\text{H}}(z) = -\frac{\rho(z)}{\epsilon_0 \epsilon_r}
\end{split}
\end{aligned}
\end{equation}
where $n$ labels bands. $\phi_{\text{H}}=-eV_{\text{H}}$ is the Hartree potential contributed by electron-electron Coulomb interactions.  The explicit form of $H_{nn^{\prime}}$ is shown in Appendix \ref{ModelHam}. $\psi_n(z)$ is $z$-dependent component of the envelope function $F_n(\bm{r})$:
\begin{equation}
F_n(\bm{r}) = e^{i(k_xx+k_yy)} \psi_n(z)
\end{equation}
where $k_x$ and $k_y$ are wave vectors. Note that they are still good quantum numbers, while $k_z$ is replaced by the differential operator $-i \frac{\partial}{\partial z}$ because of the confinement in $z$ direction. $\rho(z)$ in Eq.(\ref{Hamiltonian}) is the charge density along $z$ direction, which is discretized into $N$ pieces in numerical calculation,
\begin{widetext}
\begin{equation}\label{chargedensity}
\rho(z_j) = \frac{-e}{\Delta z} \int \frac{d^2\bm{k}}{(2\pi)^2} \bigg[ \sum\limits_{m = 1}^{8N} f(\mu-\varepsilon_m(\bm{k})) \sum\limits_{n=1}^8 \lvert \psi_n^{(m)}(z_j, \bm{k}) \rvert ^2 \bigg] - \rho_{\text{bg}}(z_j)
\end{equation}
\end{widetext}
where $f(\mu-\varepsilon_m(\bm{k}))$ is the Fermi-Dirac function and $\rho_{\text{bg}}$ is the background charge distribution calculated in the flat band condition. We used the 2D analogy of the tetrahedron method\cite{tetrahedron} when integrating over the 2D $k$-space above in Eq.(\ref{chargedensity}). We ran simulations on TACC\cite{TACC}.

A uniaxial strain opens a gap between $\Gamma_8$ light- and heavy-hole subbands and thus makes HgTe a semiconductor.
The opened band gap is proportional to the strain strength. The strain effect is added through the Bir-Pikus Hamiltonian and replacing band structure parameters with hydrostatic and uniaxial deformation potentials(more details see appendix \ref{ModelHam})\cite{Hamiltonian2005}. A 0.3\% uniaxial strain is induced by the lattice mismatch between CdTe and HgTe. The uniaxial strain tensor is given by\cite{Strain2011}:
\begin{equation}
\begin{pmatrix}
\epsilon & 0 & 0 \\
0 & \epsilon & 0 \\
0 & 0 & \frac{-2C_{12}}{C_{11}}\epsilon
\end{pmatrix}
\end{equation}
with $\epsilon=0.003$ and $ C_{12}/C_{11}=0.68$ which are chosen from \cite{Hamiltonian2005}.

\begin{figure}
\centering
\captionsetup[subfigure]{labelformat=empty}
\subfloat[]{
	\label{bandstructure}
	\includegraphics[scale=0.258]{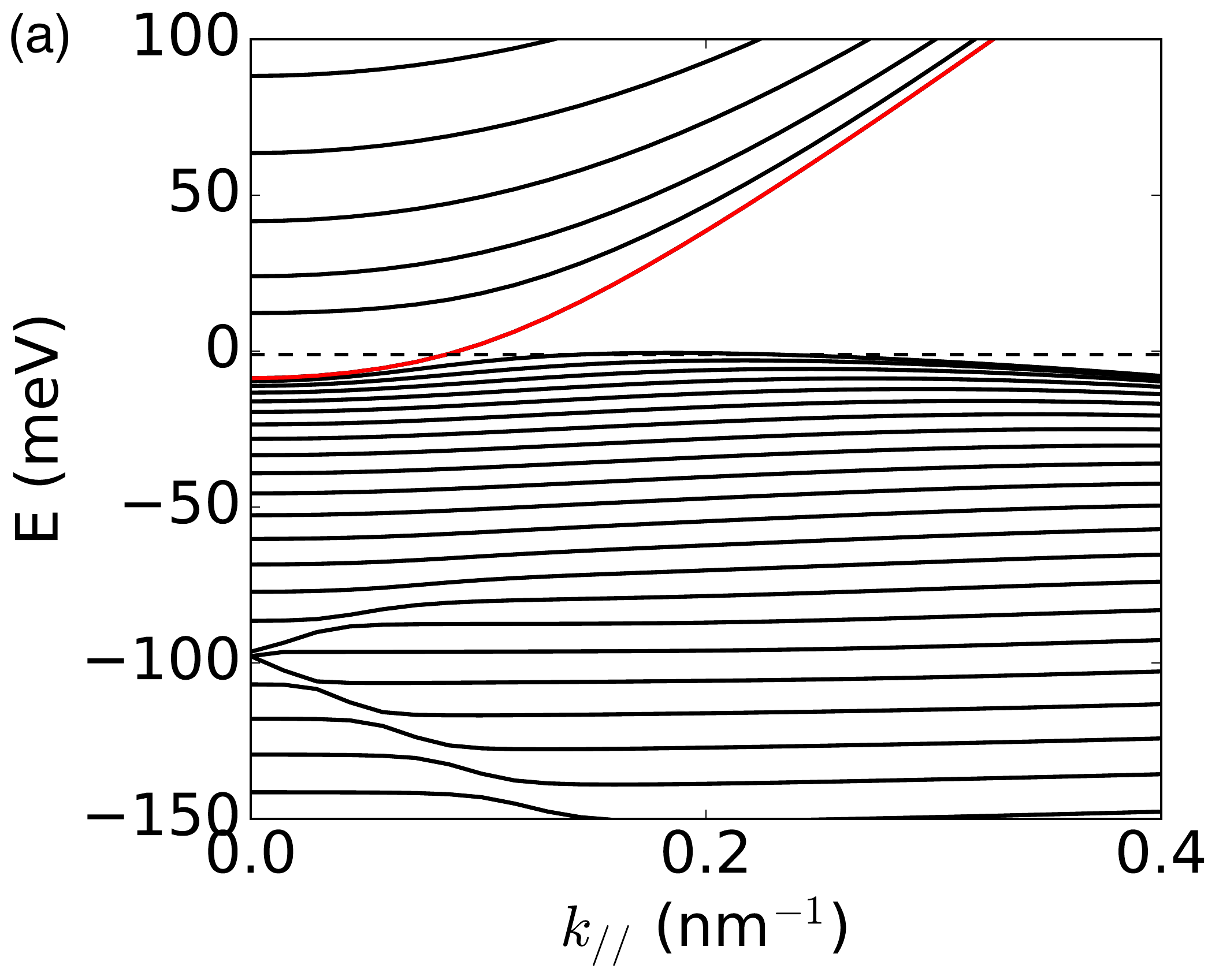}
	}
\subfloat[]{
	\label{fermisurface}
	\includegraphics[scale=0.16]{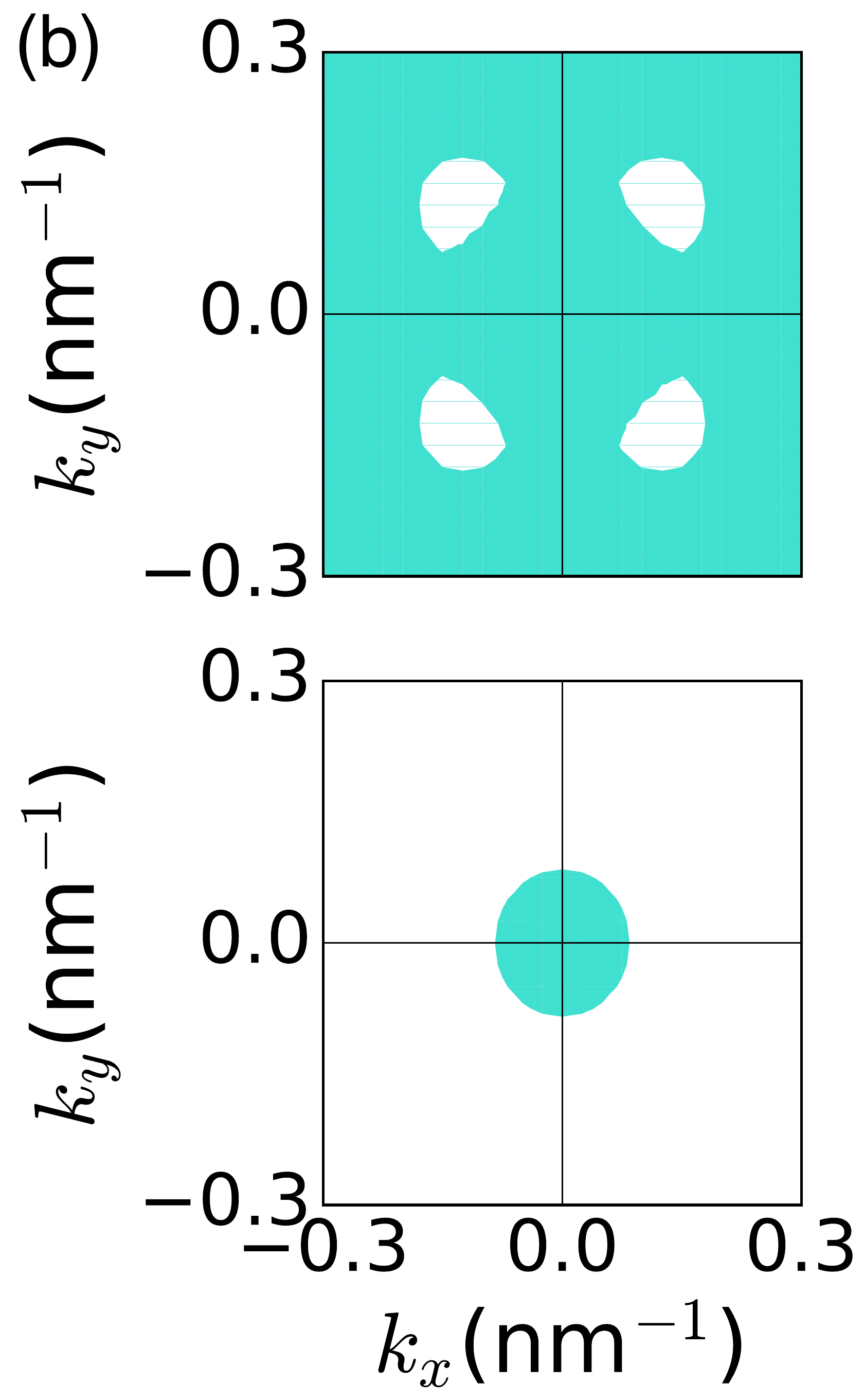}
	}
\vspace{-15pt}
\caption{
(a) The band structure of 60nm HgTe quantum well with 0.3\% uniaxial strain in the flat band condition. Black dashed line denotes the chemical potential. The red line shows the degenerate surface states.
(b) Fermi surfaces mapping onto the 2D $k$-space. Aqua region is occupied while while region is unoccupied. Upper is the top valence band and lower is the surface band.
}
\end{figure}

We show the band structure of a 60nm-thick HgTe quantum well in FIG.\ref{bandstructure}. It is under the 0.3\% uniaxial strain and is in the flat band condition.
Two degenerate surface states from top an bottom interfaces are shown with the red line in the $20$meV bulk gap.
The Dirac points are beneath the valence bands by around $100 \text{meV}$. These are consistent with former numerical results\cite{Strain2011}.

FIG.\ref{fermisurface} shows the Fermi surfaces mapping onto the 2D $k$-space in the flat band condition. It clearly shows electrons and holes coexist at charge neutrality.

\section{Static dielectric constant}\label{DielectricConstant}
The dielectric constant $\epsilon_r$ in Eq.(\ref{Hamiltonian}) is usually taken to be 21\cite{Dielectric1972}, which is the HgTe bulk dielectric constant examined by optical experiments. It includes the contributions as the following \cite{Dielectric1974}:
\begin{equation}
\epsilon_r = \epsilon_{\infty} + \epsilon_{\text{inter}} + \epsilon_{\text{intra}} \approx 21
\end{equation}
where $\epsilon_{\infty}$ is the high-frequency dielectric constant due to all interband transitions except $\Gamma_8-\Gamma_8$, $\epsilon_{\text{inter}}$ stems from $\Gamma_8-\Gamma_8$ interband transitions and $\epsilon_{\text{intra}}$ is contributed by intraband transitions in $\Gamma_8$ bands. In HgTe quantum wells, however, the static dielectric constant should be dominated by transitions among bands near the Fermi level. Hence high-frequency part $\epsilon_{\infty}=15.2$\cite{Dielectric1972} does not contribute to the dielectric constant of HgTe thin films. Meanwhile phonon contributions $\epsilon_{\text{ph}}$ should be considered. According to the effective charge\cite{Dielectric1974}, phonon dielectric contribution is estimated to be $\epsilon_{\text{ph}} \approx 0.7$. As a result, HgTe quantum well dielectric constant is $\epsilon_r \approx 6.5$, much smaller than the widely used value of 21.

When the Fermi level is inside the bulk gap, where only Dirac-like surface states are occupied, a small dielectric constant helps maintain the Fermi level in the bulk gap when increasing the carrier density. It is a result of the linear dispersion and imperfect localization of the surface states. A correction term should be added into the Fermi energy-carrier density relation due to the imperfect localization of the surface states,
\begin{equation}\label{eq:7}
\varepsilon_{\text{F}} = \hbar v_{\text{F}} \sqrt{2\pi n_e} - \lambda \frac{e^2l n_e}{\epsilon_0 \epsilon_r}
\end{equation}
where $l$ is the localization length of the surface states and $\lambda$ is a constant factor capturing the effective electric field at distance $l$ away from the top interface. Negative charges accumulate at the top interface for a positive top gate voltage. For a 50nm HgTe quantum well, $l$ and $\lambda$ can be estimated in FIG.2 to be: $l \approx 10 \text{nm}, \lambda \approx 0.7$. By the simple model illustrated in Eq.(\ref{eq:7}), energy difference of the conduction band bottom $\varepsilon_c$ and the Fermi level $\varepsilon_{\text{F}}$ as a function of carrier density $n_e$ for a 50nm HgTe is shown in FIG.3, where Fermi velocity is taken to be $v_{\text{F}} \approx 0.5 \times 10^6 \text{m}/\text{s}$\cite{fermivelocity}. Negative compressibility is only observable for $\epsilon_r<10$. \\

\begin{figure}
\centering
\captionsetup[subfigure]{labelformat=empty}
\subfloat[]{
	\label{localization1}
	\includegraphics[scale=0.23]{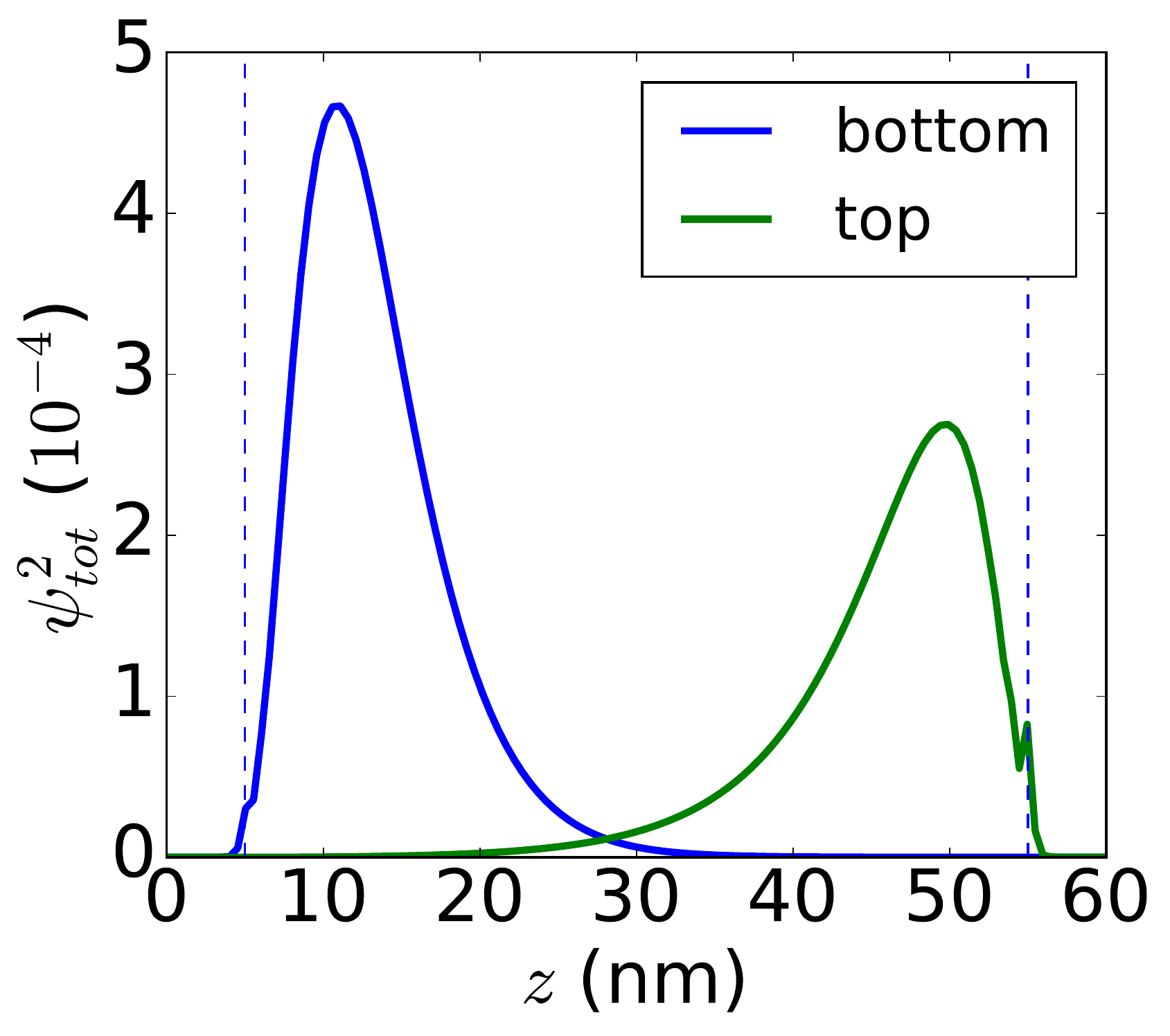}
	}
\subfloat[]{
	\label{localization2}
	\includegraphics[scale=0.23]{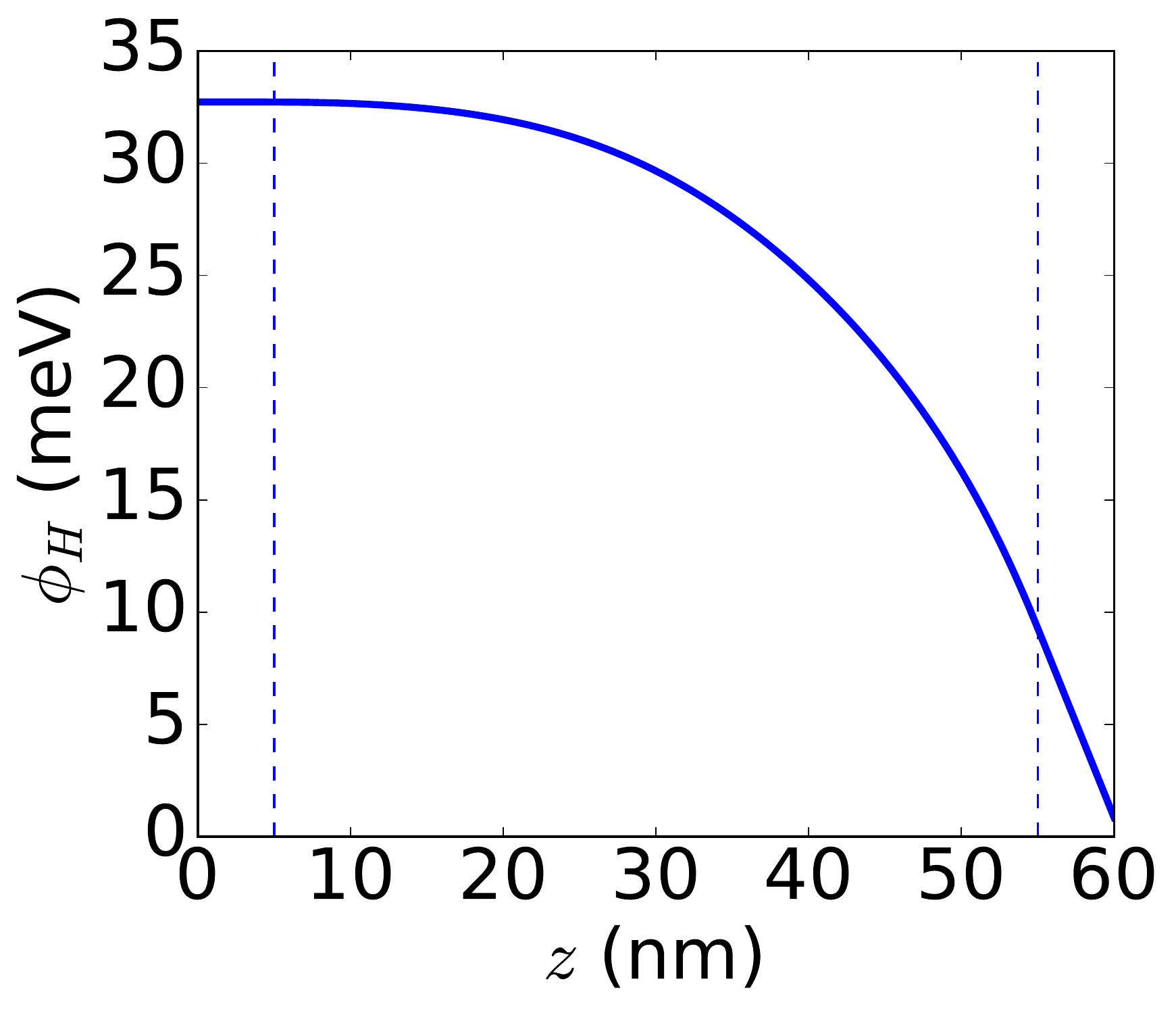}
	}
\vspace{-15pt}
\caption{
Both subfigures are based on a 50nm quantum well with $\epsilon_r = 21$, carrier density is $n_e \approx 2 \times 10^{11} \text{cm}^{-2}$. The dashed lines denote the interfaces of the quantum well and barriers.
(a) Probability density of bottom and top surfaces at the Fermi level.
(b) Hartree potential along $z$.
}
\end{figure}

\begin{figure}
\includegraphics[scale=0.4]{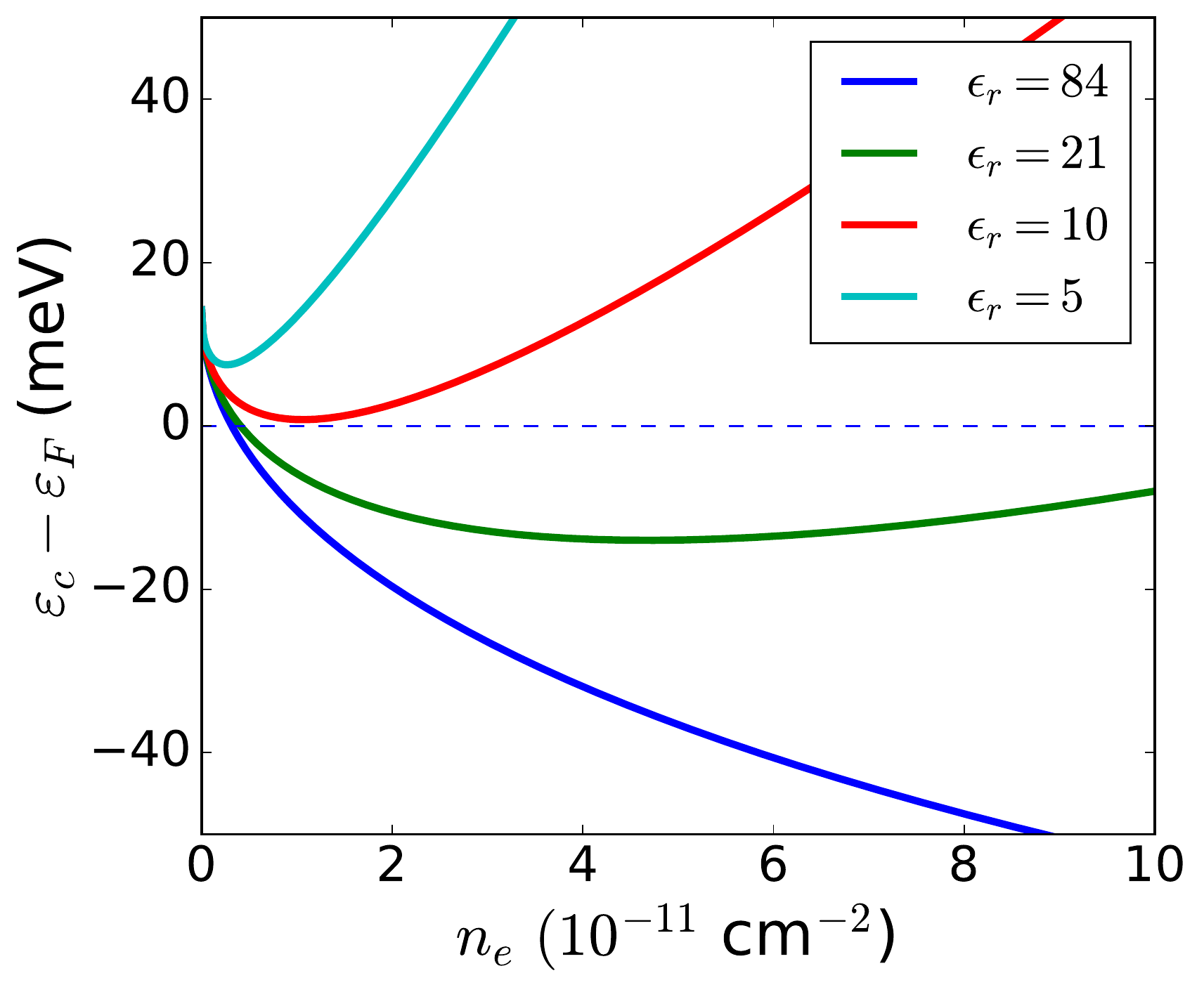}
\vspace{-10pt}
\caption{Energy difference between the conduction band bottom and the Fermi level as a function of the carrier density $n_e$ for dielectric constants $\epsilon_r=$84, 21, 10 and 5. In this simple model, the negative compressibility is only observable for $\epsilon_r < 10$.}
\end{figure}

\section{Quantum Capacitance}\label{QuantumCapacitance}
In HgTe quantum well experiments, a top gate voltage is usually applied to tune the Fermi level, thereby adjusting the carrier density in the system. In our theoretical model, we use the chemical potential $\mu$ as a parameter to adjust the carrier density in the quantum well and then convert it to the corresponding gate voltage $V_{\text{g}}(\mu)$ by:
\begin{equation} \label{eq:8}
\begin{aligned}
eV_{\text{g}} &= eV_{\text{in}} + \mu - \mu_{0} \\
&= \frac{e^2 n_e}{C_{\text{in}}} +  \mu - \mu_{0}
\end{aligned}
\end{equation}
where $V_{\text{in}}$ the electric voltage drop across the insulator between top gate and HgTe, $C_{\text{in}}$ is the geometric capacitance of the insulator and $\mu_0$ is the chemical potential in flat band condition. Take the derivative with respect to $n_e$ on both sides of Eq.(\ref{eq:8}):
\begin{equation}
\frac{1}{e}\frac{dV_\text{g}}{dn_e} = \frac{1}{C_{\text{in}}} + \frac{1}{e^2}\frac{d\mu}{dn_e}
\end{equation}
then the total capacitance $C$ can be expressed as:
\begin{equation}\label{eq:10}
\frac{1}{C} = \frac{1}{C_{\text{in}}} + \frac{1}{e^2 \frac{dn_e}{d\mu}}
\end{equation}

Total capacitances for 30-70nm HgTe quantum wells are shown in FIG.4, which is consistent with recent experimental result\cite{Qcapacitance2016}. Compare with the experiment, $e^2dn_e/d\mu$ term in Eq.(\ref{eq:10}) consists of the quantum capacitances from top ($e^2D_{\text{t}}$) and bottom ($e^2D_{\text{b}}$) interfaces and the geometric capacitance of the HgTe well ($C_{\text{tb}}$):
\begin{equation}
e^2 \frac{dn_e}{d\mu} = e^2D_{\text{t}} + \Big(\frac{1}{C_{\text{tb}}}+\frac{1}{e^2D_{\text{b}}} \Big)^{-1}
\end{equation}
where $D_{\text{t}}$ and $D_{\text{b}}$ are thermodynamic density of states of top and bottom interfaces respectively.

\begin{figure}
\centering
\captionsetup[subfigure]{labelformat=empty}
\subfloat[]{
	\label{fig:subfig_a}
	\includegraphics[scale=0.23]{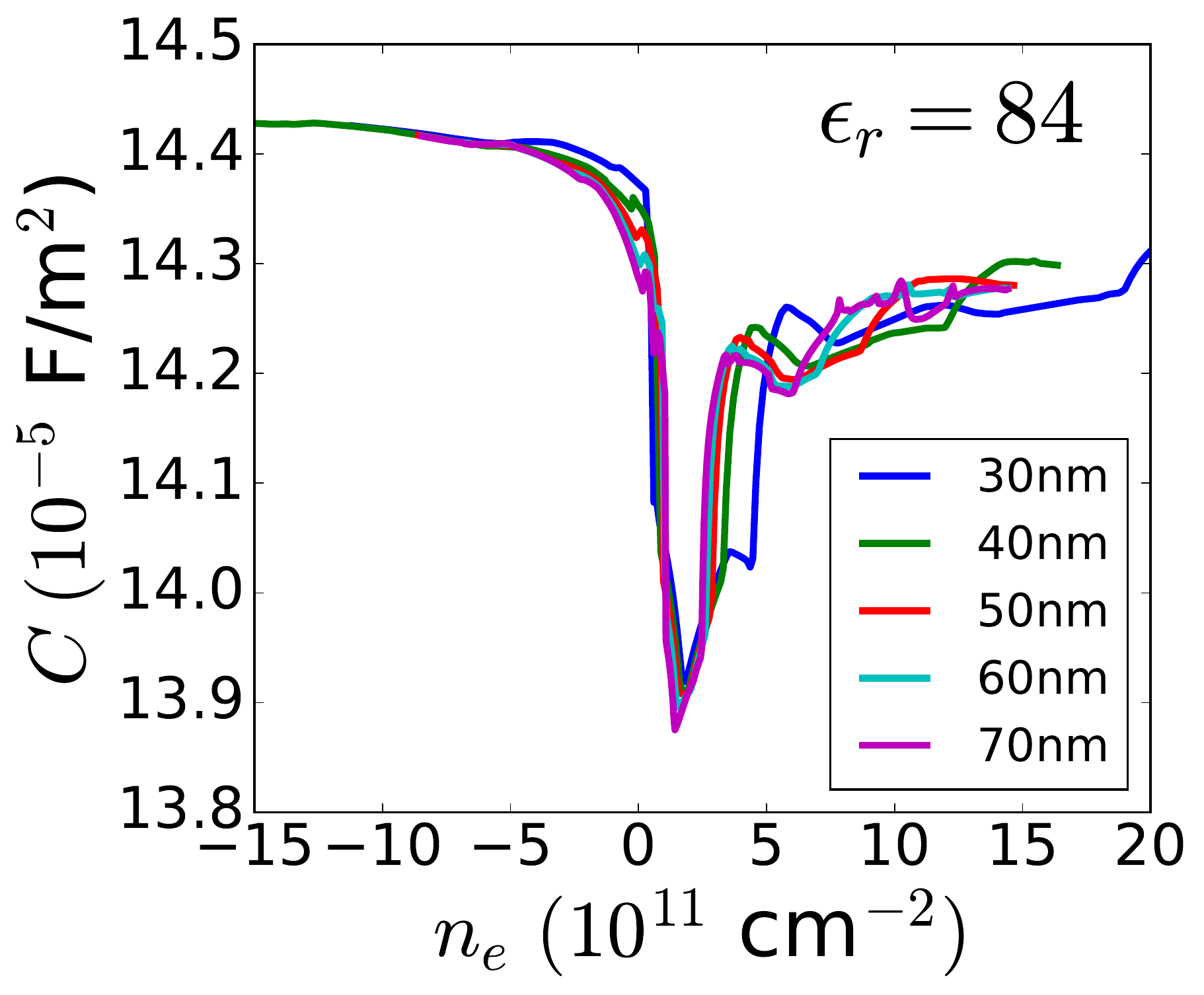}
	}
\subfloat[]{
	\label{fig:subfig_b}
	\includegraphics[scale=0.23]{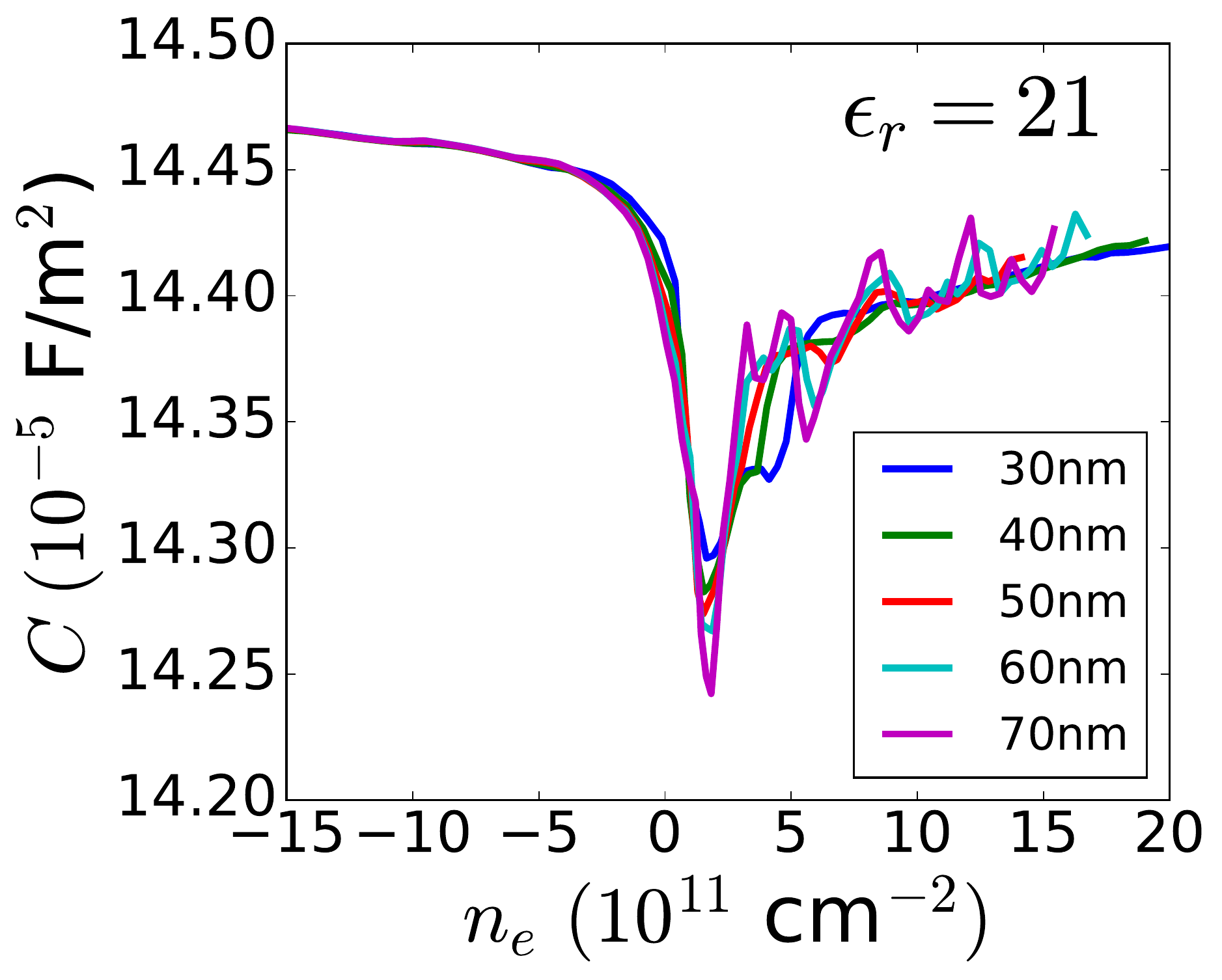}
	}\\
\vspace{-20pt}
\subfloat[]{
	\label{fig:subfig_b}
	\includegraphics[scale=0.23]{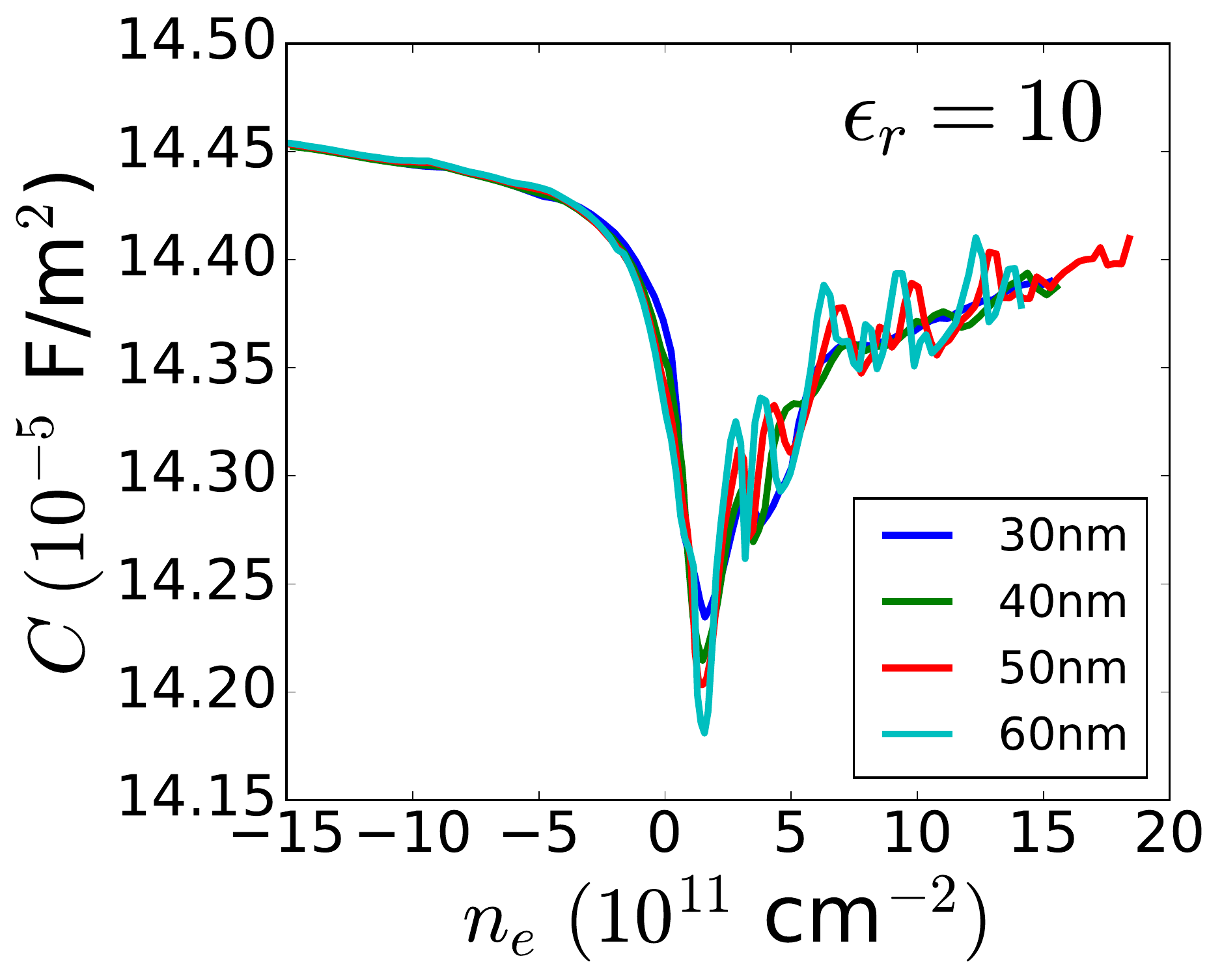}
	}
\subfloat[]{
	\label{fig:subfig_b}
	\includegraphics[scale=0.23]{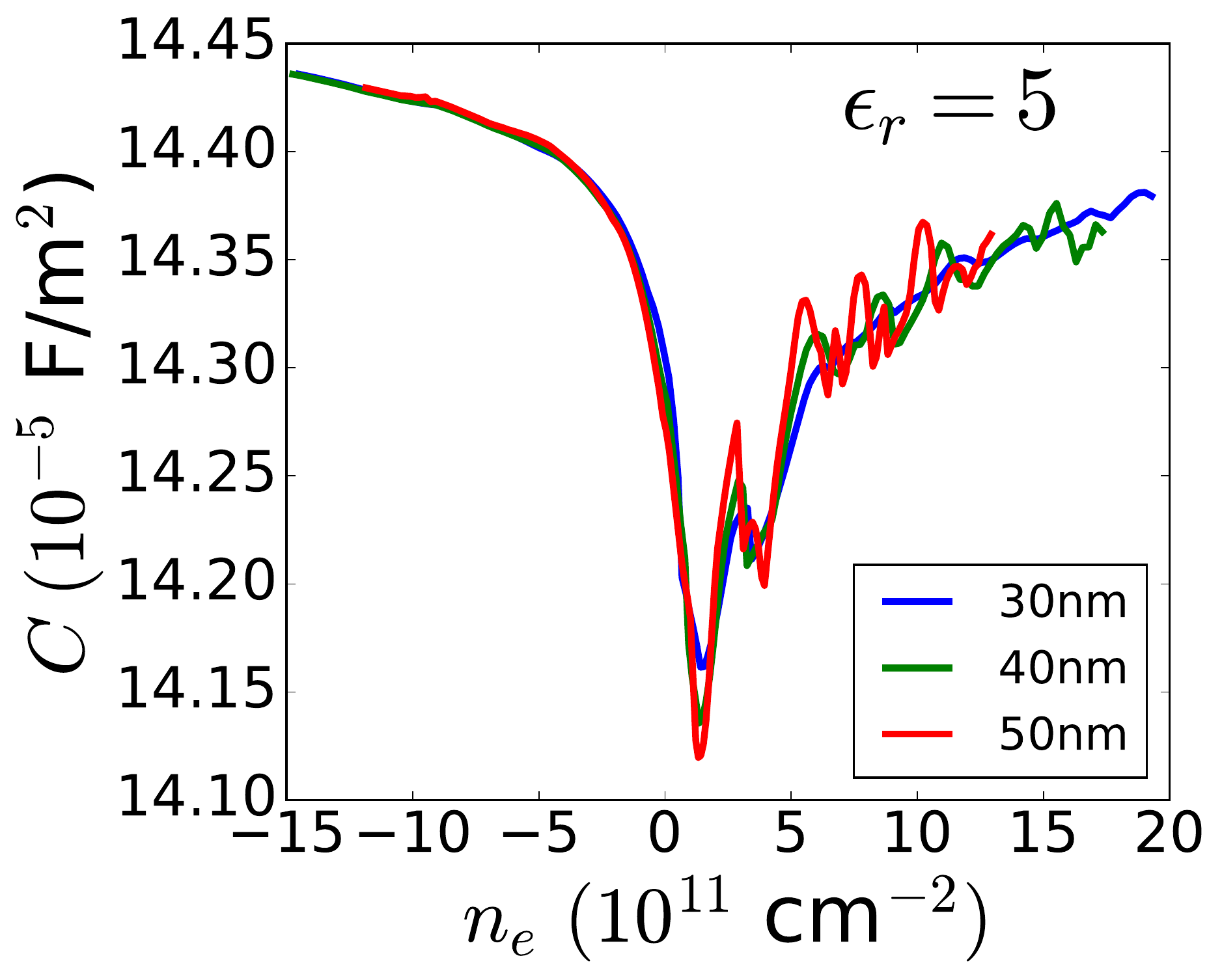}
	}
\vspace{-15pt}
\caption{
Total capacitances of 30-70nm quantum wells with dielectric constants $\epsilon_r=84,21,10, 5$. $C_{\text{in}}=1.45\times 10^{-22} \text{F}/\text{nm}^2$ is used as estimated in experiment\cite{Qcapacitance2016}.
}
\end{figure}

The quantum effect, as a result of the Pauli exclusion principle, is negligible for metals with infinitely large thermodynamic DoS. When the DoS getting smaller and ultimately comparable with the geometric capacitance, the quantum effect starts playing an important role: decreasing (increasing) the total capacitance if the thermodynamic DoS, $dn_e/d\mu$, is positive (negative). Negative thermodynamic DoS is also known as negative compressibility. As shown in FIG.4, the total capacitance reveals a depression when the chemical potential is tuned within the bulk gap, where the DoS is pretty small. The capacitance deep in the hole-dominant region is larger than the electron-dominant region. This can be simply explained by the band structure: the DoS of the valence bands is larger than that of conduction bands.

Compare the capacitances with different dielectric constants in FIG.4, the overall shape of the line is getting smoother when the dielectric constant is getting smaller. That indicates the Fermi level is resisted to be tuned into the quasi-2D conduction and valence bands. It is consistent with what we expected in section \ref{DielectricConstant}.


\section{Thermoelectric Transport}\label{Thermoelectrics}
Our theoretical band structure model can also provide convincing thermoelectric transport results within the semiclassical regime, $\omega_c\tau << 1$, here $\omega_c=\frac{eB}{m}$ is the cyclotron frequency. 

Consider the linear response to weak fields, the electric current density is expressed as:
\begin{equation}\label{j_sigma}
\bm{J} = \sigma \bm{\Sigma} + \alpha (-\bm{\nabla}T)
\end{equation}

\noindent where $\bm{\Sigma} = \bm{E}+\frac{1}{e}\bm{\nabla}\mu$ and $\mu$ is the chemical potential. Generally, $\sigma$ and $\alpha$ are tensors.
Under the relaxation time approximation and consider a uniform and spatially uniform magnetic field $\bm{B}$, the electric current density is:
\begin{widetext}
\begin{equation}\label{elecurrent}
\bm{J}=-e \int \frac{d^2 \bm{k}}{(2\pi)^2} \bigg( -\frac{\partial f^0(\varepsilon_{\bm{k}})}{\partial \varepsilon_{\bm{k}}} \bigg) \bm{v}(\bm{k})
\Big[ \tau \bm{v}(\bm{k}) + \tau^2 \frac{e}{\hbar c} (\bm{v}(\bm{k}) \times \bm{B} )_{\alpha} \frac{\partial \bm{v}(\bm{k})}{\partial k_{\alpha}} \Big]
 \cdot
 \Big[ -e\bm{\Sigma} + \frac{\varepsilon(\bm{k})-\mu}{T} \big(-\bm{\nabla}T \big) \Big]
\end{equation}
\end{widetext}
where a constant scattering time $\tau$ is assumed. Details of derivation of Eq.(\ref{elecurrent}) are in Appendix \ref{thermoelectric_appendix}. Compare Eqs.(\ref{j_sigma}) and (\ref{elecurrent}), we get expressions of conductivity $\sigma$ and coefficient $\alpha$ in terms of 2D $k$-space integration, showed in Appendix \ref{thermoelectric_appendix}. For low external field, $\omega_c \tau \ll 1$, Onsager relation $\sigma_{\mu\nu}(B) = \sigma_{\nu\mu}(-B)$ restricts the leading order term of longitudinal and transverse coefficients to be:
\begin{equation}\label{leadingorder}
\begin{aligned}
\sigma_{xx}, \alpha_{xx} &\sim \tau \\
\sigma_{xy}, \alpha_{xy} &\sim \omega_c \tau^2
\end{aligned}
\end{equation}

Consider the case that the temperature gradient is applied in $x$ direction, then the thermopower $S_{xx}$ and the Nernst coefficient $S_{xy}$ are:
\begin{equation}\label{eq14}
\begin{split}
S_{xx} = \frac{E_x}{\nabla_x T} = \frac{\alpha_{xx}\sigma_{yy}-\alpha_{yx}\sigma_{xy}}{\sigma_{xx}\sigma_{yy}-\sigma_{yx}\sigma_{xy}} \\
S_{xy} = \frac{E_y}{\nabla_x T} = \frac{\alpha_{xx}\sigma_{yx}-\alpha_{yx}\sigma_{xx}}{\sigma_{xy}\sigma_{yx}-\sigma_{yy}\sigma_{xx}}
\end{split}
\end{equation}
Inspired by Eq.(\ref{leadingorder}), we could safely ignore the terms $\sigma_{xy}\alpha_{yx} \propto (\omega_c \tau)^2$ and $\sigma_{xy}\sigma_{yx} \propto (\omega_c \tau)^2$ in Eq.(\ref{eq14}). Then,
\begin{equation}
\begin{split}
&S_{xx} \approx \frac{\alpha_{xx}}{\sigma_{xx}} \\
&S_{xy} \approx \frac{\alpha_{yx}\sigma_{xx} - \alpha_{xx}\sigma_{yx}}{\sigma_{yy}\sigma_{xx}}
\end{split}
\end{equation}
By simplified the problem with a constant scattering time and weak external field, the thermopower only depends on the band structure and the Nernst coefficient is proportional to $B\tau$ times a quantity which also only depends on the band structure.


Seebeck coefficients for 30-70nm HgTe quantum wells at a low temperature without the magnetic field are shown in FIG.5. And the Nernst effect for a 60nm quantum well is shown in FIG.6. They agree well with recent thermoelectric transport experiments\cite{Thermalpower2017}, where the Seebeck and Nernst coefficients are quite large on the hole-dominant side compared with the electron-dominant side. In our calculated results, the Seebeck and Nernst coefficients both change sign on hole- and electron-dominant sides. In the transport experiment\cite{Thermalpower2017}, the Seebeck coefficient changes sign while the Nernst coefficient keeps the same sign. That indicates that when the band structure getting complicated deep into the valence bands, the constant scattering time approximation used in our model is not a proper model to capture all effects.

\begin{figure}
\centering
\captionsetup[subfigure]{labelformat=empty}
\subfloat[]{
	\label{fig:subfig_a}
	\includegraphics[scale=0.23]{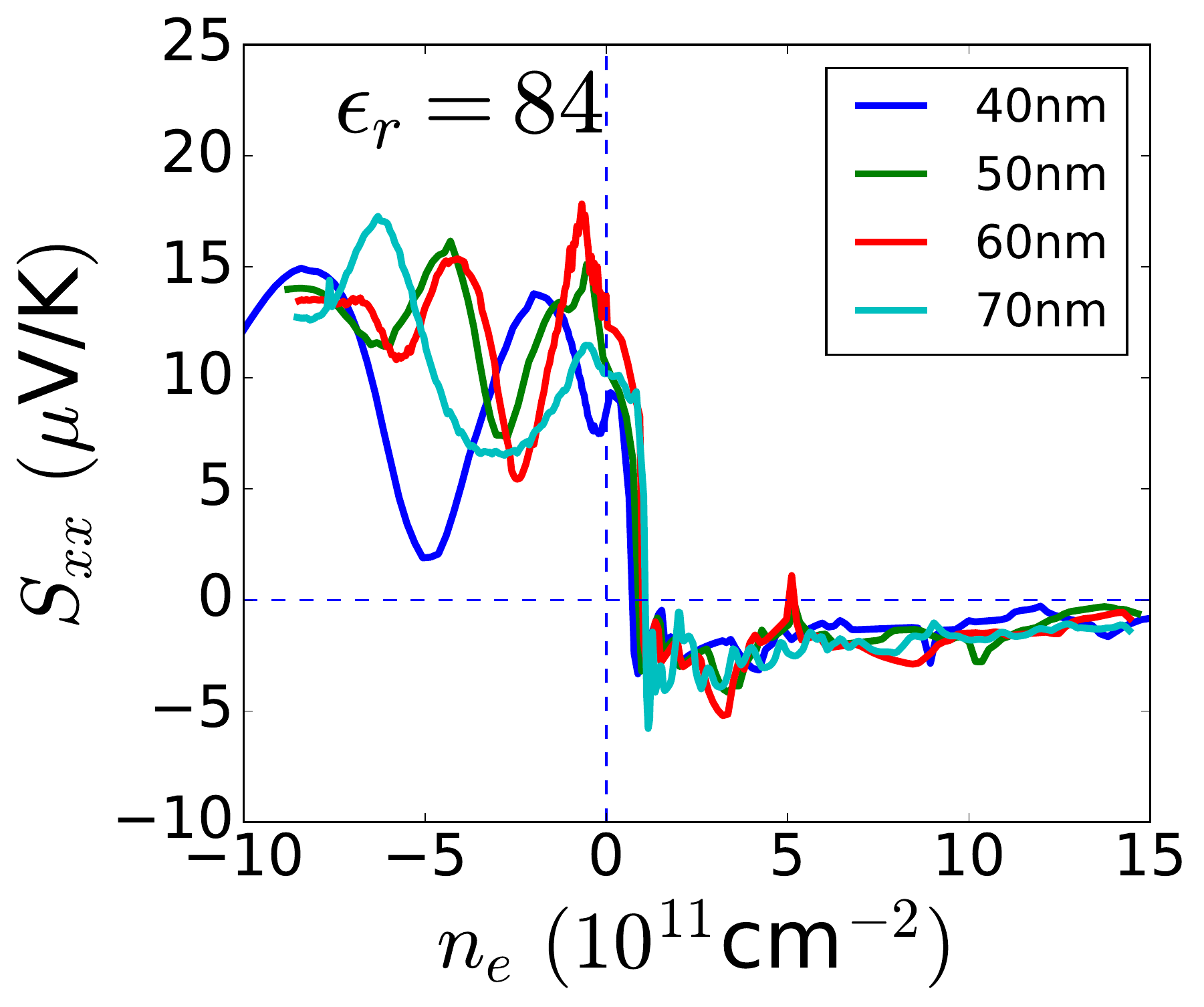}
	}
\subfloat[]{
	\label{fig:subfig_a}
	\includegraphics[scale=0.23]{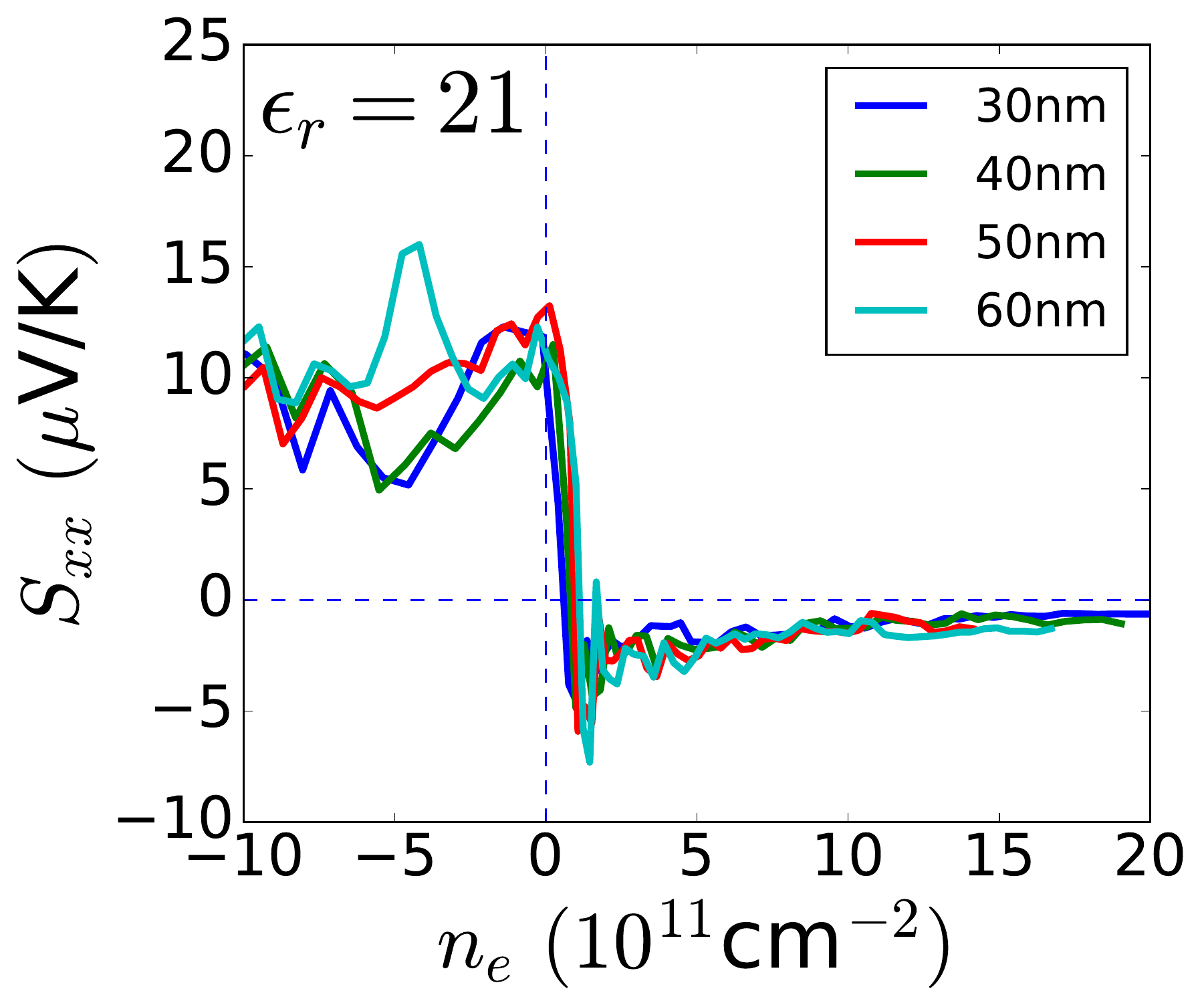}
	}
\vspace{-20pt}
\\
\subfloat[]{
	\label{fig:subfig_a}
	\includegraphics[scale=0.23]{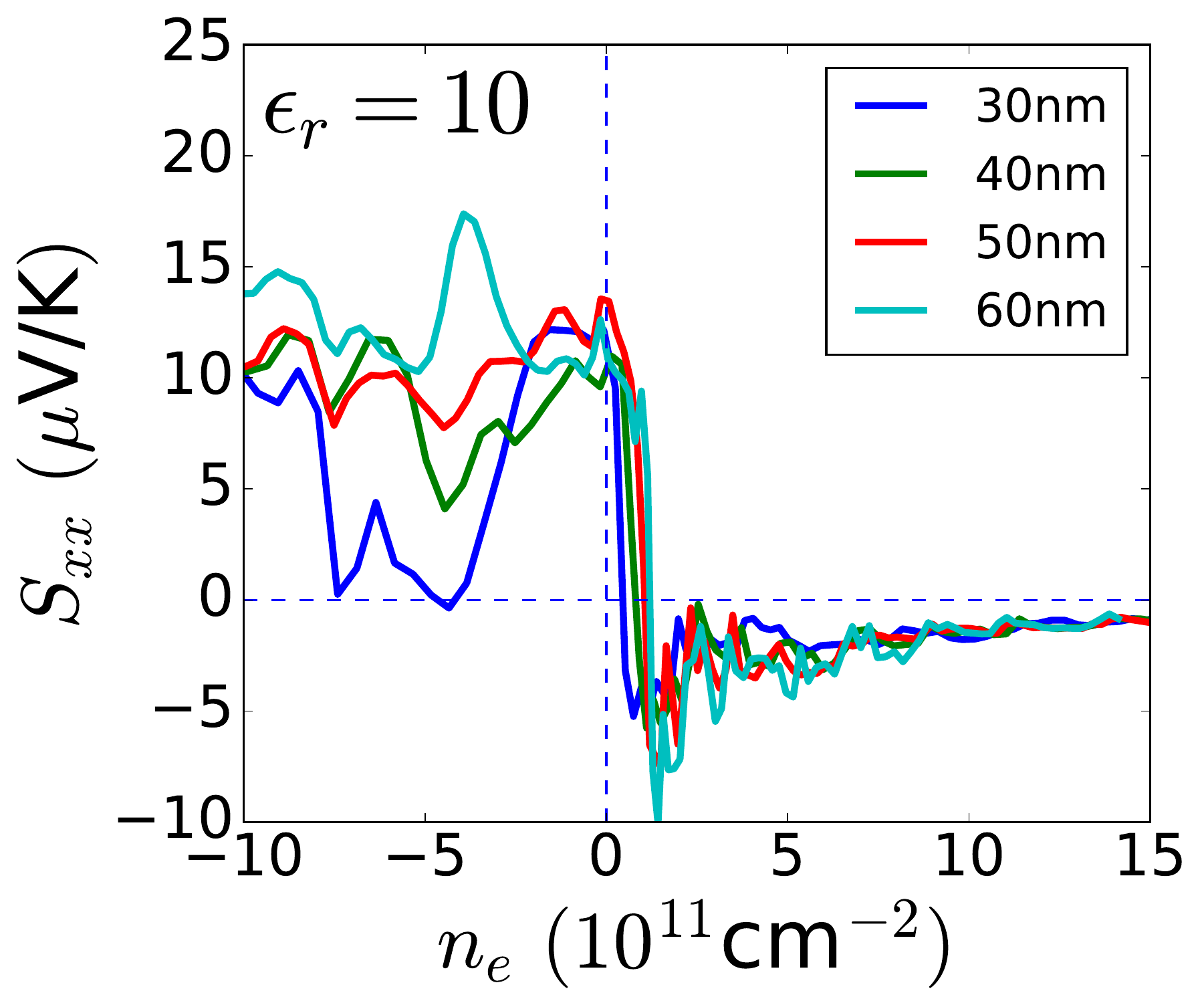}
	}
\subfloat[]{
	\label{fig:subfig_a}
	\includegraphics[scale=0.23]{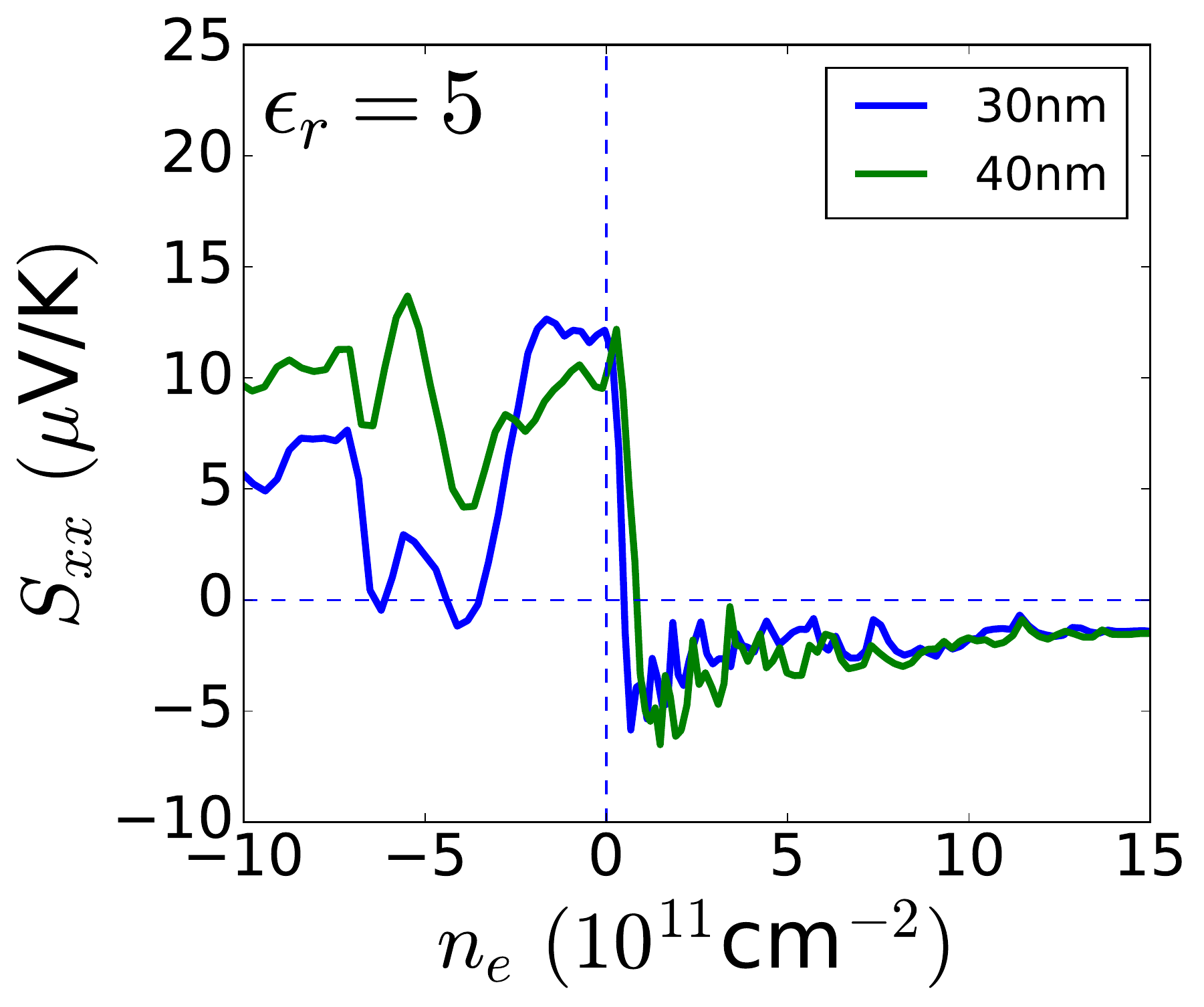}
	}
\vspace{-15pt}
\caption{
Seebeck coefficients at $B=0$ for 30-70nm HgTe quantum wells. Holes dominate on the left side of the vertical dashed line while electrons dominate on the right side. The thermopower of the hole-dominant carriers are much larger than that of electron-dominant carriers.
}
\end{figure}

\begin{figure}
\centering
\captionsetup[subfigure]{labelformat=empty}
\subfloat[]{
	\label{fig:subfig_b}
	\includegraphics[scale=0.4]{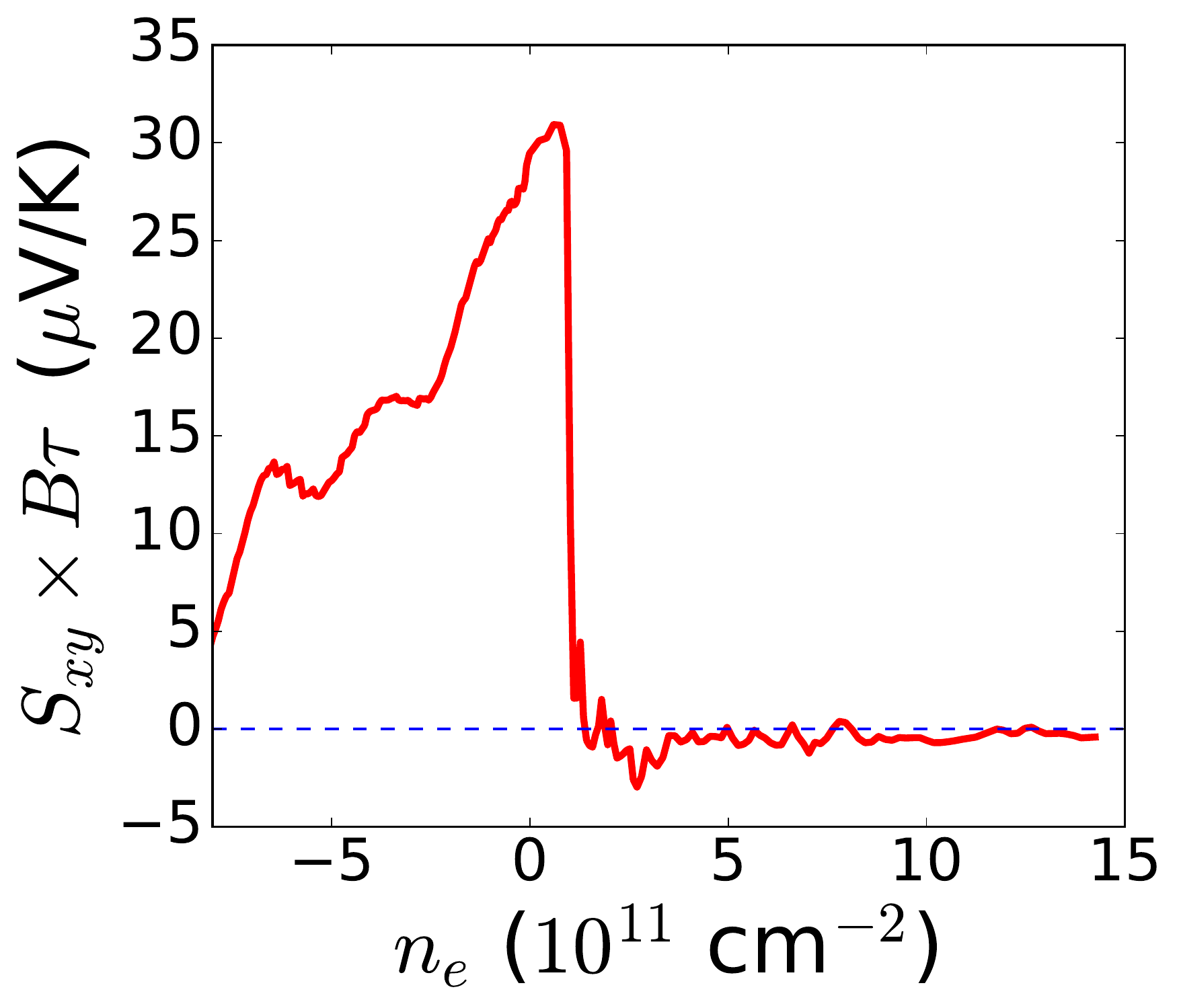}
	}
\vspace{-15pt}
\caption{
Nernst coefficient at weak magnetic field for the 60nm quantum well. $\varepsilon_r=84$. On the hole side, the Nernst coefficient is larger than the electron side. It changes sign from hole- to electron-side.
}
\end{figure}

\section{Conclusions}
According to the capacitance results both from experiments and our numerical results, the Fermi level can be easily tuned into the quasi-2D conduction and valence bands. When the Fermi level is inside the bulk gap, the maximum carrier density is around $4 \times 10^{11} \text{cm}^{-2}$ for 30-70nm quantum wells. The dielectric constant is estimated to be $\epsilon_r \approx 6$, and the small dielectric constant tries to maintain the Fermi level inside the bulk gap due to the 2D linear dispersion and the large localization length of the surface states. 

The phenomenological effective potential\cite{Diracscreening} for the sake of keeping the Fermi level within the bulk gap is not proper. Quantum Hall plateaus observed in magnetotransport studies\cite{Strain2011,Diracscreening} imply that the transport is surface-state dominated, whereas this is not the evidence that only the surface states are occupied. Inconspicuous quantized Hall resistance and non-zero longitudinal resistivity measurements $\rho_{xx}$ give a clue that quasi-2D conduction bands are also occupied, especially for high gate voltages.  

Our model shows that the thermopower is fairly large on the hole-dominated side compared to that on the electron-dominated side. The agreement with thermopower experimental results indicates that the asymmetric of the thermopower is a direct result of the band structure. The proposed \cite{Thermalpower2017} phonon drag effect to explain the thermopower asymmetry is not necessary.
By using the semiclassical transport model simplified by constant scattering time approximation, our model can capture the shape of the Nernst coefficient versus carrier density, that is, the Nernst coefficient is larger on the hole-dominated side than electron-dominated side. The sign, however, of the Nernst coefficient does not match the experiment. This indicates the constant scattering time approximation is not enough to capture all effects in this system.

\appendix
\section{Model Hamiltonian of Strained $\text{HgTe}$}\label{ModelHam}

Choose the 8-band basis set\cite{Hamiltonian2005}:
\begin{equation}
\begin{aligned}
	u_1(\mathbf{r}) & =|\Gamma_6,+\frac{1}{2} \rangle = S \uparrow \\
	u_2(\mathbf{r}) & =|\Gamma_6,-\frac{1}{2} \rangle = S \downarrow \\
	u_3(\mathbf{r}) & =|\Gamma_8,+\frac{3}{2} \rangle = \frac{1}{\sqrt{2}}(X+iY) \uparrow \\
	u_4(\mathbf{r}) & =|\Gamma_8,+\frac{1}{2} \rangle = \frac{1}{\sqrt{6}}[(X+iY) \downarrow - 2 Z \uparrow] \\
	u_5(\mathbf{r}) & =|\Gamma_8,-\frac{1}{2} \rangle = -\frac{1}{\sqrt{6}}[(X-iY) \uparrow + 2 Z \downarrow] \\
	u_6(\mathbf{r}) & =|\Gamma_8,-\frac{3}{2} \rangle = -\frac{1}{\sqrt{2}}(X-iY) \downarrow \\
	u_7(\mathbf{r}) & =|\Gamma_7,+\frac{1}{2} \rangle = \frac{1}{\sqrt{3}}[(X+iY) \downarrow + Z \uparrow] \\
	u_8(\mathbf{r}) & =|\Gamma_7,-\frac{1}{2} \rangle = \frac{1}{\sqrt{3}}[(X-iY) \uparrow - Z \downarrow] \\	
\end{aligned}
\end{equation}
The corresponding Hamiltonian of the quantum well with [001] growth direction is
\begin{widetext}
\begin{equation}
H =
\begin{pmatrix}
T  &  0   & -\frac{1}{\sqrt{2}}Pk_+ & \sqrt{\frac{2}{3}}Pk_z & \frac{1}{\sqrt{6}}Pk_-  & 0  & -\frac{1}{\sqrt{3}}Pk_z & -\frac{1}{\sqrt{3}}Pk_- \\

0  &  T   & 0  & -\frac{1}{\sqrt{6}}Pk_+ & \sqrt{\frac{2}{3}}Pk_z  & \frac{1}{\sqrt{2}}Pk_-   & -\frac{1}{\sqrt{3}}Pk_+ & \frac{1}{\sqrt{3}}Pk_z \\

-\frac{1}{\sqrt{2}}k_- P & 0 & U+V & -\bar{S}_- & R & 0 & \frac{1}{\sqrt{2}}\bar{S}_- & -\sqrt{2}R \\

\sqrt{\frac{2}{3}}k_z P & -\frac{1}{\sqrt{6}}k_- P & -\bar{S}_-^{\dagger} & U-V & C & R & \sqrt{2} V & -\sqrt{\frac{3}{2}} \widetilde{S}_- \\

\frac{1}{\sqrt{6}}k_+ P & \sqrt{\frac{2}{3}}k_z P & R^{\dagger} & C^{\dagger} & U-V & \bar{S}_+^{\dagger} & -\sqrt{\frac{3}{2}} \widetilde{S}_+ & -\sqrt{2}V \\

0 & \frac{1}{\sqrt{2}}k_+ P & 0 & R^{\dagger} & \bar{S}_+ & U+V & \sqrt{2}R^{\dagger} &  \frac{1}{\sqrt{2}} \bar{S}_+ \\

-\frac{1}{\sqrt{3}}k_z P & -\frac{1}{\sqrt{3}}k_- P & \frac{1}{\sqrt{2}} \bar{S}_-^{\dagger} & \sqrt{2} V & -\sqrt{\frac{3}{2}} \widetilde{S}_+^{\dagger} & \sqrt{2} R & U-\Delta & C \\

-\frac{1}{\sqrt{3}}k_+ P & \frac{1}{\sqrt{3}}k_z P & -\sqrt{2} R^{\dagger} & -\sqrt{\frac{3}{2}} \widetilde{S}_-^{\dagger} & -\sqrt{2}V & \frac{1}{\sqrt{2}} \bar{S}_+^{\dagger} & C^{\dagger} & U-\Delta
\end{pmatrix}
\end{equation}
where the elements in the matrix are
\begin{equation}
\begin{aligned}
& k_{\parallel}^2 = k_x^2+k_y^2,\   k_{\pm} = k_x \pm i k_y,  \ k_z = -i \partial_z \\
& T = E_c(z)+\frac{\hbar^2}{2m_0}[(2F+1) k_{\parallel}^2 + k_z(2F+1) k_z] \\
& U = E_v(z) -\frac{\hbar^2}{2m_0}(\gamma_1 k_{\parallel}^2 +k_z \gamma_1 k_z) \\
& V =-\frac{\hbar^2}{2m_0}(\gamma_2 k_{\parallel}^2 -2 k_z \gamma_2 k_z) \\
& R =-\sqrt{3} \frac{\hbar^2}{2m_0}(\mu k_+^2 -\bar{\gamma} k_-^2) = \sqrt{3} \frac{\hbar^2}{2m_0} [\gamma_2(k_x^2-k_y^2)-2i \gamma_3 k_x k_y] \\
& \bar{S}_{\pm} = -\sqrt{3} \frac{\hbar^2}{2m_0} k_{\pm} (\{ \gamma_3,k_z \} + [\kappa,k_z]) \\
& \widetilde{S}_{\pm} = -\sqrt{3} \frac{\hbar^2}{2m_0} k_{\pm} (\{ \gamma_3,k_z \} - \frac{1}{3} [\kappa,k_z]) \\
& C = 2 \frac{\hbar^2}{2m_0} k_-[\kappa,k_z] \\
\end{aligned}
\end{equation}
\end{widetext}

The effects of strain are added to the Hamiltonian through the Bir-Pikus Hamiltonian\cite{Strain2011,Strain2014}:
\begin{equation}
H_{n n^{\prime}} \rightarrow H_{n n^{\prime}}+H_{n n^{\prime}}^{BP}
\end{equation}
$H^{BP}$ is derived from the Hamiltonian without strain by substitution:
\begin{equation}
k_i k_j \rightarrow \varepsilon_{ij}
\end{equation}
and the band structure parameters are replaced by:
\begin{equation}
\begin{split}
&\frac{\hbar^2}{2m_0}(2F+1) \rightarrow a_c \\
&\frac{\hbar^2}{m_0} \gamma_1 \rightarrow -2a_v \\
&\frac{\hbar^2}{m_0} \gamma_2 \rightarrow -b \\
&\frac{\hbar^2}{m_0} \gamma_3 \rightarrow -d/\sqrt{3}
\end{split}
\end{equation}
where\cite{Strain2011}
\begin{equation}
\begin{split}
& a_c = -4.6 \text{eV},\\
& a_v=-0.13\text{eV}, \\
& b = -1.15\text{eV}, \\
& C_{12}/C_{11}=0.68
\end{split}
\end{equation}
$a_c$ and $a_v$ are the hydrostatic deformation potentials of the conduction and valence bands, respectively, and $b, d$ are uniaxial deformation potentials.

\section{Thermoelectric Effects Under Weak Fields}\label{thermoelectric_appendix}
Seebeck effect is the longitudinal thermoelectric voltage induced by a temperature gradient, and Nernst effect is the transverse thermoelectric response of a temperature gradient under a out-of-plane magnetic field.
Microscopic theory of these thermoelectric effects in the semiclassical picture can be described with the Boltzmann equation in steady state:
\begin{equation}\label{BoltzmannEq}
 \bm{v}(\bm{k}) \cdot \bm{\nabla}_{\bm{r}}f^0_{\bm{k}} + \frac{\bm{F}}{\hbar} \cdot \bm{\nabla}_{\bm{k}}f^0_{\bm{k}} = \Big( \frac{\partial f}{\partial t} \Big)_{\text{coll}}
\end{equation}

\noindent where $f^0_{\bm{k}} = \big( e^{(\varepsilon_{\bm{k}}-\mu)/k_{\text{B}}T}+1 \big)^{-1}$ is the Fermi-Dirac distribution function describing the equilibrium state, $f=f(\bm{r},\bm{k},t)$ is the distribution function deviating from the equilibrium, $ \bm{F}$ is the external forces induced by electric field, magnetic field or temperature gradient. 
The right hand side of Eq.(\ref{BoltzmannEq}) is the collision effect. Here we consider the relaxation time approximation for simplicity:
\begin{equation}\label{relaxationApproximation}
\Big( \frac{\partial f}{\partial t} \Big)_{\text{coll}} = -\frac{1}{\tau} \big(f(\bm{r},\bm{k},t)-f^0_{\bm{k}} \big)
\end{equation}
where a constant relaxation time $\tau $ is assumed.

The solution of Eq.(\ref{BoltzmannEq}) is\cite{ashcroft1976}:
\begin{equation}
 f_{\bm{k}}(t) - f^0_{\bm{k}} =  - \int_{-\infty}^{t} {dt^{\prime} P(t,t^{\prime}) \frac{d f^0_{\bm{k}} }{dt^{\prime}}}
\end{equation}
where $ P(t,t^{\prime}) = e^{-(t-t^{\prime})/\tau} $ is the fraction of electrons survive from $t^{\prime}$ to $ t $ without collisions. With semiclassical equations of motion:
\begin{equation}
\begin{split}
&\dot{\bm{r}} =\bm{v}(\bm{k}) = \frac{1}{\hbar} \frac{\partial \varepsilon_{\bm{k}}}{\partial \bm{k}} \\
&\hbar \dot{\bm{k}} = -e \big[ \bm{E}(\bm{r},t)+\frac{1}{c}\bm{v}(\bm{k}) \times \bm{B}(\bm{r},t) \big]
\end{split}
\end{equation}
and use the fact that:
\begin{equation}
\begin{split}
 &\frac{d f^0_{\bm{k}} }{dt} = \frac{d \bm{k}}{dt} \cdot \bm{\nabla}_{\bm{k}} f_{\bm{k}}^{0} 
+  \frac{d \bm{r}}{dt} \cdot \bm{\nabla}_{\bm{r}} f_{\bm{k}}^{0} \\
& \frac{1}{\hbar} \bm{\nabla}_{\bm{k}} f_{\bm{k}}^{0} = \bm{v}({\bm{k}}) \frac{\partial f_{\bm{k}}^0}{\partial \varepsilon_{\bm{k}}} \\
&  \bm{\nabla}_{\bm{r}} f_{\bm{k}}^{0} = \frac{\partial f_{\bm{k}}^{0} }{\partial T} \bm{\nabla}_{\bm{r}}T + \frac{\partial f_{\bm{k}}^{0} }{\partial \mu} \bm{\nabla}_{\bm{r}}\mu
\end{split}
\end{equation}
then
\begin{equation}\label{B6}
\begin{split}
 f_{\bm{k}}(t)- f^0_{\bm{k}}=& \int_{-\infty}^{t} dt^{\prime} e^{-(t-t^{\prime})/\tau}  \Big( -\frac{\partial f_{\bm{k}}^0 }{\partial \varepsilon_{\bm{k}}} \Big) \bm{v}\big( \bm{k}(t^{\prime}) \big) \cdot \\
&\Big[ -e\bm{E}(t^{\prime}) -\bm{\nabla}_{\bm{r}}\mu(t^{\prime})-\frac{\varepsilon_{\bm{k}}-\mu}{T} \bm{\nabla}_{\bm{r}}T  \Big]
 \end{split}
\end{equation}
The magnetic field is out of the Eq.(\ref{B6}) if it is perpendicular to the transport plane, $\bm{B} = B\hat{z}$.

The deviation function $\phi_{\bm{k}}$, defined by $
f_{\bm{k}} \approx f^0_{\bm{k}} - \frac{\partial f^0_{\bm{k}}}{\partial \varepsilon_{\bm{k}}} \phi_{\bm{k}}
$, is directly derived from Eq.(\ref{B6}):
\begin{equation}
\begin{split}
\phi_{\bm{k}}(t) = & \int_0^{\infty} dt^{\prime \prime} e^{-t^{\prime \prime}/\tau} \bm{v}\big( \bm{k}-t^{\prime \prime} \dot{\bm{k}}\big) \cdot  \Big[ -e\bm{E}(t-t^{\prime \prime}) \\
& -\bm{\nabla}_{\bm{r}} \mu (t-t^{\prime \prime}) - \frac{\varepsilon_{\bm{k}}-\mu}{T} \bm{\nabla}_{\bm{r}} T(t-t^{\prime \prime})\Big]
\end{split}
\end{equation}
where $t^{\prime \prime} = t-t^{\prime}$. Ignore the time dependence of $\bm{E}$, $\mu$ and $T$, and use
\begin{equation}
\begin{split}
&\bm{k}(t-t^{\prime \prime}) = \bm{k}(t)-t^{\prime \prime}\dot{\bm{k}} \\
&\bm{v}(\bm{k}-t^{\prime \prime} \dot{\bm{k}}) = \bm{v}(\bm{k})-t^{\prime \prime} \dot{\bm{k}} \cdot \frac{\partial \bm{v}}{\partial \bm{k}} 
\end{split}
\end{equation}

\noindent then to the first order in fields,
\begin{equation}
\begin{split}
\phi_{\bm{k}} =& \Big[ \tau \bm{v}({\bm{k}})+\tau^2 \frac{eB}{\hbar c} \big(\bm{v}({\bm{k}}) \times \hat{z} \big)_{\alpha} \frac{\partial \bm{v}({\bm{k}})}{\partial k_{\alpha}} \Big] \cdot \\
& \Big[ -e\bm{\Sigma} + \frac{\varepsilon_{\bm{k}}-\mu}{T}(-\bm{\nabla} T) \Big]
\end{split}
\end{equation}

\noindent where $\bm{\Sigma} = \bm{E}+\frac{1}{e} \bm{\nabla} \mu$. Here we ignore the spatial subscript in the gradient label $\bm{\nabla}$, and subscript $\alpha$ is the Einstein notation. 

The electric current density is expressed as
\begin{equation}
\begin{aligned}
\bm{J} &= -e\int \frac{d^2\bm{k}}{(2\pi)^2} \  \bm{v}({\bm{k}}) \Big( -\frac{\partial f^0_{\bm{k}}}{\partial \varepsilon_{\bm{k}}} \Big) \phi_{\bm{k}} \\
&=  -e\int \frac{d^2\bm{k}}{(2\pi)^2} \  \bm{v}({\bm{k}}) \Big( -\frac{\partial f^0_{\bm{k}}}{\partial \varepsilon_{\bm{k}}} \Big) \Big[ \tau \bm{v}({\bm{k}})+\\
&\ \ \ \ \ \ \ \ \tau^2 \frac{eB}{\hbar c} \big(\bm{v}({\bm{k}}) \times \hat{z} \big)_{\alpha} \frac{\partial \bm{v}({\bm{k}})}{\partial k_{\alpha}} \Big] \cdot \Big[ -e\bm{\Sigma} + \\
&\ \ \ \ \ \ \ \ \frac{\varepsilon_{\bm{k}}-\mu}{T}(-\bm{\nabla} T) \Big] \\
&=\sigma \bm{\Sigma} + \alpha (-\bm{\nabla} T)
\end{aligned}
\end{equation}
where
\begin{equation}
\begin{split}
\sigma = e^2 \int & \frac{d^2\bm{k}}{(2\pi)^2}  \Big( -\frac{\partial f^0_{\bm{k}}}{\partial \varepsilon_{\bm{k}}} \Big) \bm{v}({\bm{k}}) \\
&\Big[ \tau \bm{v}({\bm{k}})+ \tau^2 \frac{eB}{\hbar c} \big(\bm{v}({\bm{k}}) \times \hat{z} \big)_{\alpha} \frac{\partial \bm{v}({\bm{k}})}{\partial k_{\alpha}} \Big]
\end{split}
\end{equation}
\begin{equation}
\begin{split}
\alpha = \frac{-e}{T} \int & \frac{d^2\bm{k}}{(2\pi)^2}  \Big( -\frac{\partial f^0_{\bm{k}}}{\partial \varepsilon_{\bm{k}}} \Big) (\varepsilon_{\bm{k}}-\mu) \bm{v}({\bm{k}}) \\
&\Big[ \tau \bm{v}({\bm{k}})+ \tau^2 \frac{eB}{\hbar c} \big(\bm{v}({\bm{k}}) \times \hat{z} \big)_{\alpha} \frac{\partial \bm{v}({\bm{k}})}{\partial k_{\alpha}} \Big]
\end{split}
\end{equation}
$\sigma$ and $\alpha$ both satisfy the Onsager relation $\sigma_{\mu\nu}(B) = \sigma_{\nu\mu}(-B)$. To the leading order in magnetic field, the Onsager relation leads to:
\begin{equation}
\begin{split}
&\sigma_{xx}, \sigma_{yy}, \alpha_{xx}, \alpha_{yy} \sim \tau \\
&\sigma_{xy}, \sigma_{yx}, \alpha_{xy}, \alpha_{yx} \sim \omega_c \tau^2 \\
\end{split}
\end{equation}
where $\omega_c = \frac{eB}{m}$ is the cyclotron frequency. Specifically,
\begin{equation}\label{B14}
\begin{split}
&\sigma_{xx} = e^2\tau \int \frac{d^2 \bm{k}}{(2\pi)^2} \Big( -\frac{\partial f^0_{\bm{k}}}{\partial \varepsilon_{\bm{k}}} \Big) v_x^2 \\
& \alpha_{xx} = \frac{-e\tau}{T} \int \frac{d^2 \bm{k}}{(2\pi)^2} \Big( -\frac{\partial f^0_{\bm{k}}}{\partial \varepsilon_{\bm{k}}} \Big)(\varepsilon_{\bm{k}}-\mu) v_x^2 \\
&\sigma_{xy} = \tau^2 \frac{e^3B}{\hbar c} \int \frac{d^2 \bm{k}}{(2\pi)^2} \\
&\ \ \ \ \ \ \ \ \ \ \Big( -\frac{\partial f^0_{\bm{k}}}{\partial \varepsilon_{\bm{k}}} \Big) v_x \Big(v_y \frac{\partial v_y}{\partial k_x} - v_x \frac{\partial v_y}{\partial k_y} \Big) \\
&\alpha_{xy} = -\frac{\tau^2}{T} \frac{e^2B}{\hbar c} \int \frac{d^2 \bm{k}}{(2\pi)^2} \\
&\ \ \ \ \ \ \ \ \ \ \Big( -\frac{\partial f^0_{\bm{k}}}{\partial \varepsilon_{\bm{k}}} \Big) (\varepsilon_{\bm{k}}-\mu) v_x \Big(v_y \frac{\partial v_y}{\partial k_x} - v_x \frac{\partial v_y}{\partial k_y} \Big)
\end{split}
\end{equation}

Consider the case that the temperature gradient is applied in $x$ direction. The thermopower $S_{xx}$ and Nernst effect coeficient $S_{xy}$ are defined as
\begin{equation}\label{B15}
\begin{split}
S_{xx} = \frac{E_x}{\nabla_x T} = \frac{\alpha_{xx}\sigma_{yy}-\alpha_{yx}\sigma_{xy}}{\sigma_{xx}\sigma_{yy}-\sigma_{yx}\sigma_{xy}} \\
S_{xy} = \frac{E_y}{\nabla_x T} = \frac{\alpha_{xx}\sigma_{yx}-\alpha_{yx}\sigma_{xx}}{\sigma_{xy}\sigma_{yx}-\sigma_{yy}\sigma_{xx}}
\end{split}
\end{equation}

We can safely ignore the magnetic field's high-order terms in Eq.(\ref{B15}), then
\begin{equation}
\begin{split}
&S_{xx} \approx \frac{\alpha_{xx}}{\sigma_{xx}} \\
&S_{xy} \approx \frac{\alpha_{yx}\sigma_{xx} - \alpha_{xx}\sigma_{yx}}{\sigma_{yy}\sigma_{xx}}
\end{split}
\end{equation}
and $\sigma, \alpha$ are expressed in Eq.(\ref{B14}).

\begin{acknowledgments}
The authors would like to thank B. Xiong, C. Xiao, M. Xie, F. Xue and T. Lovorn for helpful discussions. This work was supported by the Army Research Office (ARO) under contract W911NF-15-1-0561:P00001, and by the Welch Foundation under grant TBF1473. The authors acknowledge the Texas Advanced Computing Center (TACC) at The University of Texas at Austin for providing HPC, visualization, database and grid resources that have contributed to the research results reported within this paper. \\
\end{acknowledgments}

\bibliographystyle{apsrev4-1}
\bibliography{HgTe}

\end{document}